# Tuning a sign of magnetoelectric coupling in paramagnetic $NH_2(CH_3)_2Al_{1-x}Cr_x(SO_4)_2 \times 6H_2O$ crystals by metal ion substitution

V. Kapustianyk[1,2], Yu. Eliyashevskyy[2], Z. Czapla[3,4], V. Rudyk[1], R. Serkiz[1], N. Ostapenko[2], I. Hirnyk[1], J.-F. Dayen[5], M. Bobnar[6], R. Gumeniuk[7] & B. Kundys[5,+]

Hybrid organometallic systems offer a wide range of functionalities, including magnetoelectric (ME) interactions. However, the ability to design on-demand ME coupling remains challenging despite a variety of host-guest configurations and ME phases coexistence possibilities. Here, we report the effect of metal-ion substitution on the magnetic and electric properties in the paramagnetic ferroelectric $NH_2(CH_3)_2Al_{1-x}Cr_x(SO_4)_2 \times 6H_2O$. Doing so we are able to induce and even tune a sign of the ME interactions, in the paramagnetic ferroelectric (FE) state. Both studied samples with $x = 0.065$ and $x = 0.2$ become paramagnetic, contrary to the initial diamagnetic compound. Due to the isomorphous substitution with Cr the ferroelectric phase transition temperature ($T_c$) increases nonlinearly, with the shift being larger for the 6.5% of Cr. A magnetic field applied along the polar $c$ axis increases ferroelectricity for the $x = 0.065$ sample and shifts $T_c$ to higher values, while inverse effects are observed for $x = 0.2$. The ME coupling coefficient $\alpha_{ME} = 1.7$ ns/m found for a crystal with Cr content of $x = 0.2$ is among the highest reported up to now. The observed sign change of $\alpha_{ME}$ with a small change in Cr content paves the way for ME coupling engineering.

Realization of the effective interactions between magnetic moments and electric charges constitutes an important task for modern solid state physics[1,2] and spin electronics[3–6]. The principal motivation is to establish electric control of magnetism for low-power spintronic structures. For this reason large values of magnetization and electric polarization are often motivating factors for multiferroic materials research[7–9]. Although large magnetization is expected from ferromagnetic ordering, ferromagnetism and ferroelectricity tend to be mutually exclusive in a single phase[10] and the largest magnetoelectric coupling is mostly seen in antiferromagnets at symmetry-breaking spin reorientation transitions[11]. However, ME coupling of higher orders can be symmetry independent and exists for other types of magnetic orderings[12,13]. In particular, a large ME effect was reported to exist in the paramagnetic $[(CH_3)_2NH_2]Mn(HCOO)_3$[14]. Along with other successful examples[15–20], this result demonstrates a large potential of organic-inorganic materials[21] in the research of ME compounds and beyond[22,23]. Because electric order in the lattice is more fragile than a magnetic one, a promising strategy to achieve their safe coexistence can be implementation of magnetic interactions into already known electrically polar compounds rather than vice versa. In this respect the organic-inorganic hybrid frameworks offer indeed an abundance of possibilities[24–32]. Here we study the ferroelectric $NH_2(CH_3)_2Al(SO_4)_2 \times 6H_2O$ (DMAAS) crystal which belong to organic-inorganic functional materials known to be electrically polar below 152K[33,34]. The crystal structure of DMAAS is built up of Al cations coordinated by six water molecules (i.e. water octahedra), regular $(SO_4)^{2-}$ tetrahedra and $[NH_2(CH_3)_2]+$ (DMA) cations, all hydrogen bonded to a three dimensional framework (Fig. 1). Thus, in the crystal structures of [NH$_2$(CH$_3$)$_2$]Al$_{1-x}$Cr$_x$(SO$_4$)$_2$ × 6H$_2$O with $x = 0.2$ and 0.065 octahedra are occupied by statistical mixtures of magnetic Cr$^-$ and nonmagnetic Al$^-$ atoms. With cooling this crystal exhibits a second order phase transition at $T_c = 152$ K from paraelectric but ferroelastic ($T > T_c$) to ferroelectric ($T < T_c$) phases. The phase transition is of the order-disorder type with a symmetry change 2/m → m. It is connected with ordering of the polar DMA cations which execute hindered rotations around their C-C direction in the paraelectric phase and order only in the spatio-temporal average in the ferroelectric phase[35]. Metal ion isomorphous substitution in the above mentioned family of compounds can be an additional degree of freedom in the composition-property engineering[36,37]. Here we report that the incorporated magnetic Cr cations can participate in ME interactions in the [NH$_2$(CH$_3$)$_2$]Al$_{1-x}$Cr$_x$(SO$_4$)$_2$ × 6H$_2$O crystals with ability to tune magneto-electric functionality.

[1]Scientific-Technical and Educational Center of Low-Temperature Studies, Ivan Franko National University of Lviv, Dragomanova str. 50, 79005, Lviv, Ukraine. [2]Department of Physics, Ivan Franko National University of Lviv, Dragomanova str. 50, 79005, Lviv, Ukraine. [3]Department of Physics, Opole University of Technology, Ozimska 75, 45370, Opole, Poland. [4]Institute of Experimental Physics, University of Wrocław, pl. M. Borna 9, 50204, Wroclaw, Poland. [5]Institute de Physique et de Chimie des Matériaux de Strasbourg, UMR 7504 CNRS-ULP, 23 rue du Loess, BP 43, F67034, Strasbourg, Cedex 2, France. [6]Max Plack Institut für Chemische Physik fester Stoffe, Nöthnitzer Str. 40, 01187, Dresden, Germany. [7]Institut für Experimentelle Physik, TU Bergakademie Freiberg, Leipziger Str. 23, 09596, Freiberg, Germany. Correspondence and requests for materials should be addressed to B.K. (+email: kundysATipcms.unistra.fr)

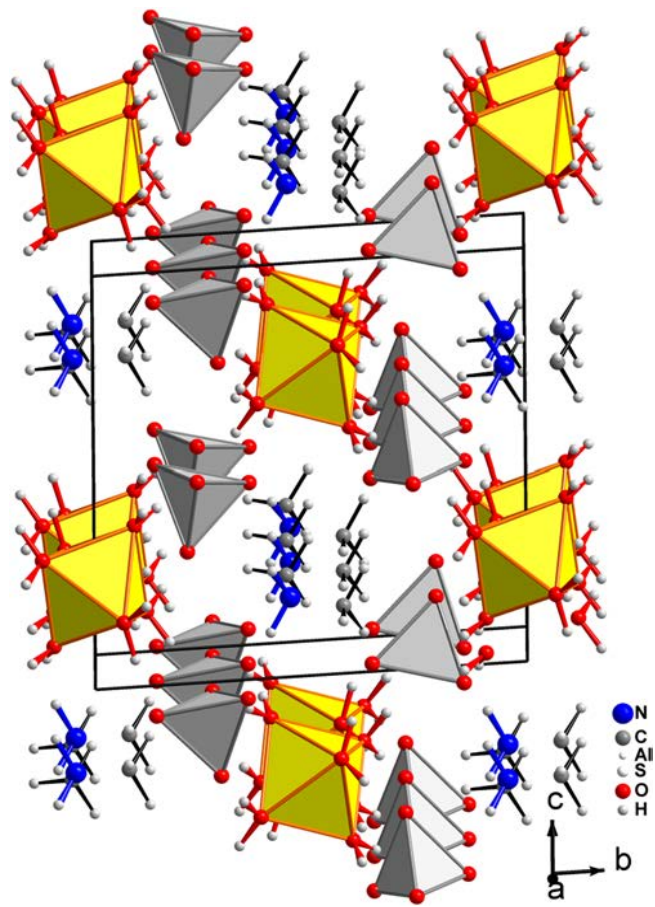

**Figure 1.** Crystal structure of non-centrosymmetric monoclinic [NH$_2$(CH$_3$)$_2$]Al(SO$_4$)$_2$ × 6H$_2$O at 135 K. The Al atoms are in the centers of yellow [H$_2$O]$_6$-octahedra. In the case of Cr substitution the octahedra are occupied by statistical mixtures of magnetic Cr and nonmagnetic Al atoms. The [SO$_4$]$^{2-}$ tetrahedra are depicted in grey color.

## Results

Initial DMAAS crystals usually grow in a monodomain ferroelastic state. The ferroelastic domains may nevertheless appear during polishing or other type of mechanical treating. However, if Cr is introduced this tendency is inversed: the samples predominately grow in a polydomain ferroelastic state (Fig. 2). In this case the EDX analysis of the neighboring domains shows a different content of chromium. The quantity of Cr in oppositely stressed domains is equal to 17.6% and 20% respectively. This rule, however, does not apply to minor quantity of samples yet grown in the single domain state where chromium is distributed evenly within a sample. It is observed that thermodynamical conditions of the growth are highly stress-dependent and allow variation of the Cr distribution within the sample volume. This intriguing result, however, deserves separate investigations that are beyond the scope of this paper. During our further discussion we will consider only the crystals grown in a polydomain states as this type of growth is statistically overwhelming. The average thickness of domains was found to be equal to 100–120 μm, with the optical indicatrix disorientation angle of 2θ = 40.

The existence of a polydomain ferroelastic state was also verified by scanning electron microscopy imaging (Fig. 2c,d). Similarly to the initial compound our samples show large peaks in the dielectric permittivity due to the polar state formation at $T_c = 153$ K which is characteristic for proper ferroelectric phase transitions (Fig. 3). Notably, the value of the dielectric permittivity is almost three orders of magnitude larger for an AC electric field applied along the $c$ axis confirming the spontaneous polarization direction (Fig. 3a). The temperature dependences of the thermal expansion measured along the three principal axes of a DMAAl$_{0.8}$Cr$_{0.2}$S crystal are presented in Fig. 3b. The clear continuous changes of all three lattice parameters are characteristic of a second order phase transition and confirm its structural component. The pyroelectric measurements for DMAAl$_{0.8}$Cr$_{0.2}$S confirm and compliment the electrically polar character of the transition.

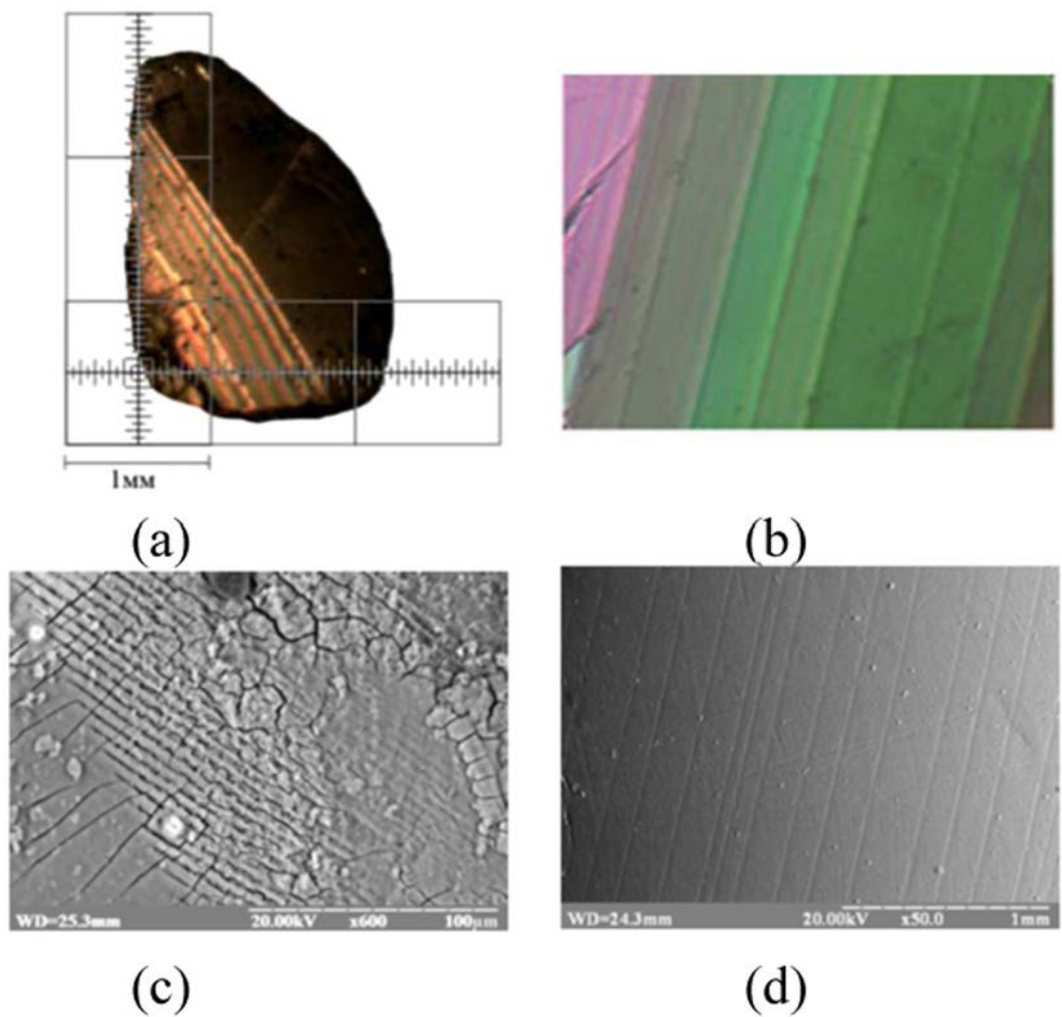

**Figure 2.** The ferroelastic domain structure at 300 K. Polarization microscopy pictures on the cut of DMAAl$_{0.8}$Cr$_{0.2}$S crystal perpendicular to [310] (**a**,**b**) and corresponding view of the (001) (**c**) and (310) (**d**) surfaces of the same crystals obtained using a scanning electron microscopy in COMPO and TOPO regimes respectively.

A distinct clear peak in the current is observed at the same temperature where both dielectric permittivity and thermal expansion also show anomalies. A DC electric field of 7.9 kV/m applied during cooling reveals peaks in the pyroelectric currents for both samples and demonstrates Cr dependent $T_c$ evolution. The temperature dependences of the dielectric properties for both investigated crystals fairly well correlate with the data of previous investigations[37] and obey the Curie-Weiss law both in the paraelectric (+) and ferroelectric (−) phases in the vicinity of the ferroelectric phase transition. Together with the data of DSC study[37] this clearly confirms a second order of the phase transition. No other anomalies were observed in the pyro-current temperature dependences at cooling of both samples down to 1.6 K. Therefore, one can conclude that the ferroelectric phase exists in our DMAAl$_{1-x}$Cr$_x$S crystals in the temperature range from $T_c$ down to 1.6 K. This conclusion is also confirmed by the temperature dependences of the electric polarization (Fig. 4a) measured after ferroelectric saturation occurring above 250 kV/m[38]. The magnetic susceptibilities of DMAAl$_{1-x}$Cr$_x$S complexes are depicted in the inset to Fig. 4b along with electric polarization data (Fig. 4a). The pure complex without Cr ($x = 0$) is diamagnetic in the whole studied temperature range with residual susceptibility $\chi_0$ given in Table 1.

Isomorphous substitution of Al with Cr leads to the appearance of a paramagnetic fraction below 100 K for $x = 0.06$ and to a paramagnetic behaviour and, thus positive $\chi(T)$ for $x = 0.2$ (inset to Fig. 4b). Both these susceptibilities fit excellently to modified Curie-Weiss (CW) law ($\chi = C/T + \chi_0$) in the temperature range 50–300 K (inset to Fig. 4b). As one can see from Table 1 the $\chi_0$ values obtained from the fit agree well with those of initial DMAAS crystal. This confirms that Cr-atoms are embedded into a diamagnetic matrix. Effective magnetic moments $\mu_{eff}$ deduced from the fit are close, indicating no Cr-Cr interactions in the studied compounds. As it is known $Cr^{3+}$ usually forms octahedral complexes. The $Al^{3+}$ ions (in this crystallographic position the Cr substitution is expected) center octahedral voids in the studied structures (Fig. 1). Both these facts hint toward $+3$ oxidation state for Cr (i.e. $3d^3$ electronic configuration) and thus, a low spin type of complex in agreement with earlier studies[36]. Interestingly, no anomalies are seen in the magnetic susceptibilities at ferroelectric $T_c$ as it would be expected from ref.[14]. However, careful examination of the derivatives $d\chi/dT$ reveals a clear deviation from linearity near $T_c$ (Fig. 4b) for the $DMAAl_{1-x}Cr_xS$ crystals and temperature independent linear behavior for the initial sample. Even more, the magnetic susceptibility derivative for the crystal with $x=0.065$ shows an upturn towards low temperatures while for $x=0.2$ an upturn occurs towards high temperatures.

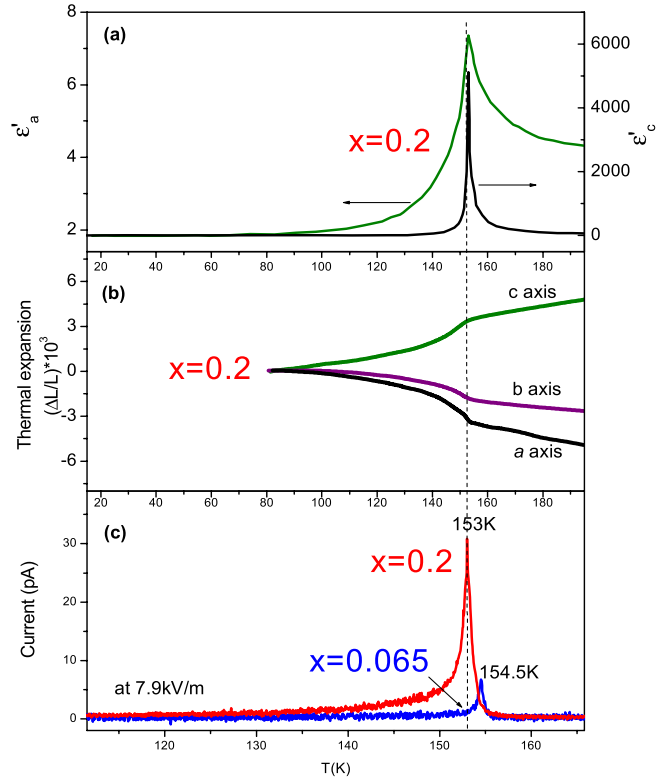

**Figure 3.** Temperature dependence of the structurally correlated electric properties. (**a**) The real part of the dielectric permittivity $\varepsilon'_a$ and $\varepsilon'_c$ for $DMAAl_{0.8}Cr_{0.2}S$ crystals. (**b**) The thermal expansion measured along the principal cuts of $DMAAl_{0.8}Cr_{0.2}S$ crystals. (**c**) Pyroelectric currents for samples with $x=0.2$ and $0.065$ Cr content.

Such a different behavior correlates well with the different magnetic field dependence of the pyroelectric current magnitude and its temperature position for both samples (Fig. 5). With increasing of magnetic field the intensity of the peak in pyro-current increases and shifts towards higher temperatures (promotes ferroelectricity) for the compound with $x=0.065$, while for $x=0.2$ the opposite effects are observed (Fig. 5). The ME coupling coefficient $\alpha_{ME}$ in the units of [s/m] is defined here as:

$$\alpha_{ME} = \frac{1}{\Delta H} \int (I_{H=O} - I_H)dt$$

where $I$ is the pyroelectric current density and $H$ is the magnetic field applied ($H\mu_0 = 6\,T$) during the measurements (Fig. 5). As one can see from the ME coupling coefficient presented in Fig. 6, the effect of coupling is stronger for the crystal with Cr content $x=0.2$. As expected the absolute value of $\alpha_{ME}$ decreases monotonically till the FE Curie temperature is reached.

However, both crystals also show a discontinuity in ME coupling near $T_C$ (inset to Fig. 6) implying the existence of a small coupling even in the paraelectric and paramagnetic region. The fact that the ME coupling coefficient changes its sign as a function of Cr content as well as the large level of the coupling itself points towards the possibility to tune the ME response in such compounds.

The ME properties can be tentatively explained by the twofold effect. Firstly, the introduction of larger ($Ri_{Cr} = 0.6115\,Å$; $Ri_{Al} = 0.535\,Å$)[39] and magnetic Cr generates strains and secondly, makes the compound more sensitive to magnetic field. The magnetoelectric coupling here can arise via stress mediated contribution. With Cr content increase the overall sample deformation goes via critical point modifying local magnetism and polarization. This assumption seems to be in agreement with the sensitive ferroelastic structure. This issue, however, deserves a separate study including optimal Cr content determination.

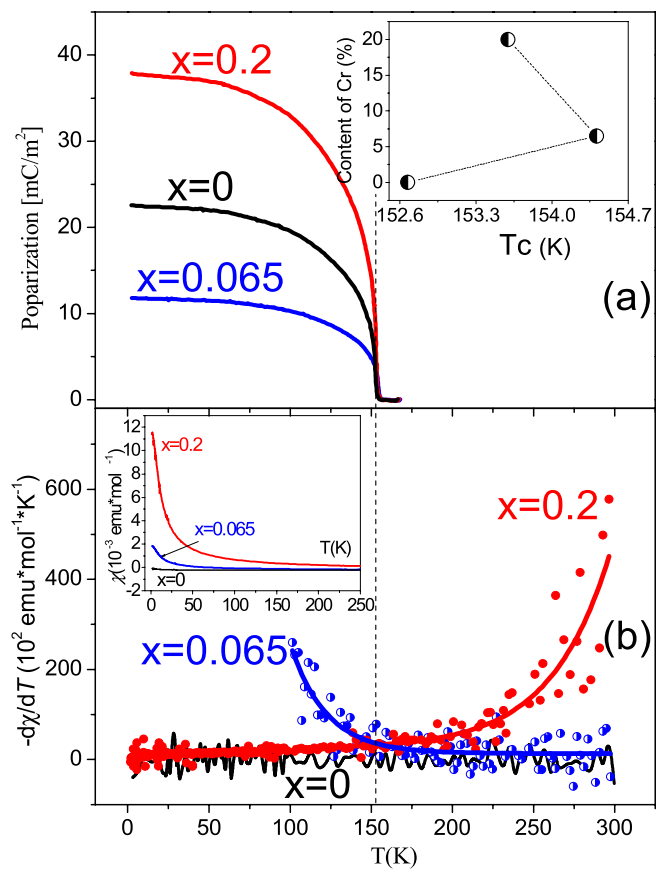

**Figure 4.** Temperature dependence of magnetic and electric properties. (**a**) Electric polarization. (**b**) The derivatives of susceptibilities for DMAAl$_{1-x}$Cr$_x$S crystals with different Cr content. Insets to figures (**a**) and (**b**) show respectively a variation of the ferroelectric temperature $T_c$ and magnetic susceptibilities from which magnetic parameters are determined via CW fit (Table 1).

| Cr-content, $x$ | $\chi_0$ ($10^{-4}$ emu mol$^{-1}$) | $\mu_{eff}$ ($\mu_B$) |
|---|---|---|
| 0 | −2.26(7) | |
| 0.06 | −2.40(5) | 1.47(1) |
| 0.2 | −2.49(8) | 1.91(1) |

**Table 1.** Magnetic parameters for DMAAl$_{1-x}$Cr$_x$S crystals.

In conclusion, this study reports the successful creation of paramagnetic order in the initially diamagnetic DMAAS crystal by isomorphous substitution of metal ion. Such action intimately connects apparently separated magnetic and electric subsystems, as evidenced by the temperature dependence of the derivative magnetic susceptibility. We have successfully generated a large ME coupling and importantly demonstrated the possibility to tune its sign, depending on the Cr content. Moreover, our results suggest that ME coupling can exist in the paramagnetic compounds without easily noticeable magnetic anomalies near FE transitions, and special care should be taken for such evidence. From the ferroelectric point of view, we show that partial isomorphous substitution with Cr metal ions leads to an anoticeable shift of the phase transition and can be used to increase FE polarization and $T_c$. In particular, in comparison with initial DMAAS crystal, the phase transition temperature $T_c$ in the crystal doped with Cr$^{3+}$ (6.5%) is shifted toward higher temperatures by 2.6 K, whereas for a higher chromium concentration (20%), this shift is diminished to 0.6 K. The Cr distribution in such samples can also be controlled by stress-assisted material growth conditions and provides additional degrees of freedom for organometallic materials engineering. Our study motivates further investigations in the area of paramagnetic organic-inorganic materials, with the design of ME interactions at room temperature as the next milestone.

## Methods

Single crystals of [NH$_2$(CH$_3$)$_2$]Al$_{1-x}$Cr$_x$(SO$_4$)$_2$ × 6H$_2$O (DMAAl$_{1-x}$Cr$_x$S) were grown from an aqueous solution containing the metal sulphates in a stoichiometric ratio and dimethylammonium sulfate at a constant temperature of 303 K by slow evaporation method. The molar ratio of Al$^{3+}$: Cr$^{3+}$ in the solution was equal to 1: 0.065 and

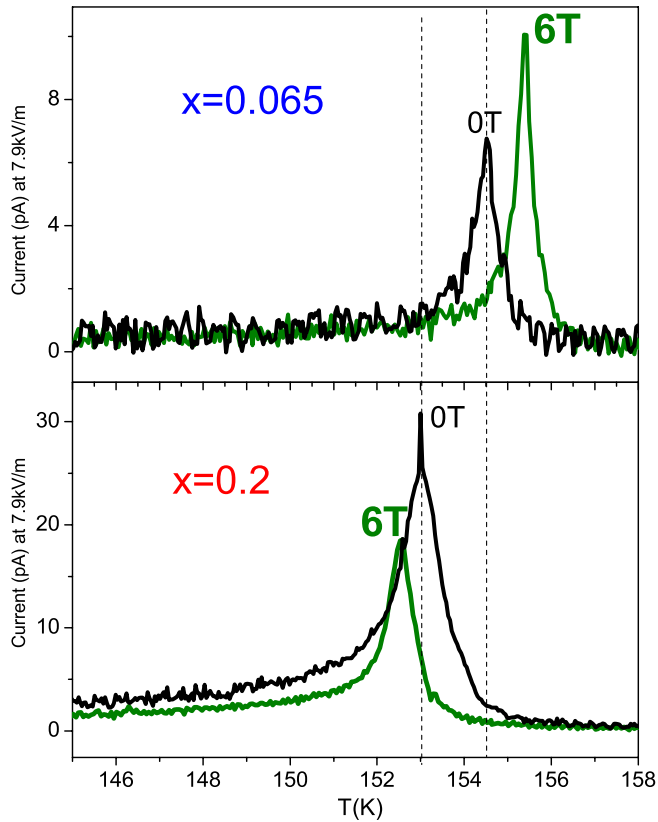

**Figure 5.** Magnetic field influence on the pyroelectricity. Upper panel: the sample with the Cr content of $x = 0.065$; Lower panel: the sample with the Cr content of $x = 0.2$. An opposite behavior in both magnitude and temperature position of the pyroelectric peak is observed.

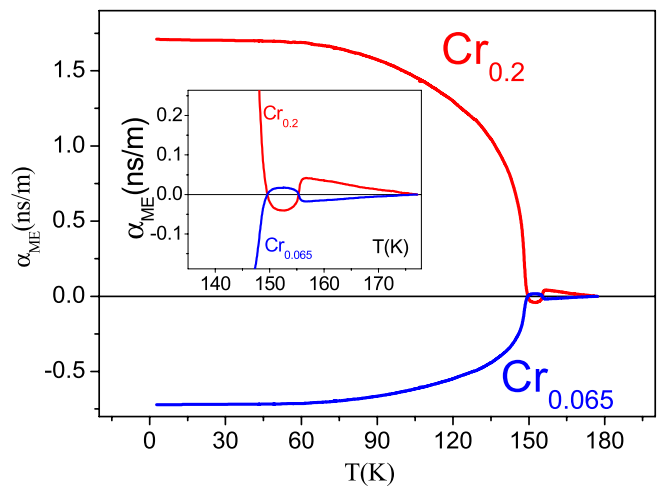

**Figure 6.** Temperature dependence of the ME coupling coefficients for parallel orientation of magnetic and electric fields. Inset shows a zoomed region near $T_c$.

1: 0.2, respectively. This ratio in the samples was controlled by SEM using a REMMA-102–02 (SELMI, Ukraine) scanning electron microscope. Quantitative electron probe microanalysis (EPMA) of the phases was carried out using an energy-dispersive X-ray (EDX) analyzer with the pure elements as standards (the acceleration voltage was 20 kV; *K*- and *L*-lines were used). The obtained values of $Al^{3+}$ : $Cr^{3+}$ molar ratio were found to be $0.065 \pm 0.006$ and $0.2 \pm 0.02$ (for a single domain samples) and correspond to those in the reacting solution. The surface morphology was studied by SEM. The scanning of sample surface was performed by an electron beam with energy of 15 and 20 kV and a diameter of 5 nm in the secondary electron image regime. To prevent charging during SEM cycling, the sample was covered by a thin graphite layer transparent for the electron beam. The thermal expansion was measured using a home-built capacitive dilatometer. The measurements of the real part of dielectric permittivity and conductivity were carried out using the traditional method of capacitor capacitance measurement. The capacitance was measured using an automated setup based on a LCR-meter HIOKI 3522-50 LCF HiTester. The spontaneous polarization measured using Keithley 6517 A electrometer. The magnetic susceptibility was measured using a commercial magnetometer Quantum Design MPMS-3 in the temperature range 1.8–300 K and magnetic fields up to $\mu_0 H = 7$ T. For both polarization and magnetic measurements electric and magnetic fields were applied perpendicular to the crystallographic plane in the monoclinic crystal structure (parallel to the polar axis).

### Acknowledgements

This work was partially supported by HYMN, hvSTRICTSPIN and LABEX research projects. Authors thank Andreas Leithe-Jasper for his careful and critical reading of this manuscript. The technical help of Fabien Chevrier is also gratefully acknowledged.